\documentclass{iau} 
\usepackage{graphicx}

\title[Do AGN really suppress star formation?] 
{Do AGN really suppress star formation?}

\author[C.M. Harrison et~al.]   
{C.M. Harrison$^{1}$, D.M. Alexander$^2$,  D.J. Rosario$^{2}$, J. Scholtz$^{2,1,3}$ \\\and
                    F. Stanley$^{3}$ 
}

\affiliation{$^1$European Southern Observatory, Karl-Schwarzschild-Str. 2, 85748
   Garching b. M{\"u}nchen, Germany \\ email: {\tt c.m.harrison@mail.com} \\[\affilskip]
$^2$Centre for Extragalactic Astronomy, Durham University, Department
of Physics, South Road, Durham, DH1 3LE, United Kindgom\\[\affilskip]
$^3$Department of Space, Earth and Environment, Chalmers University of Technology, Onsala Space Observatory, SE-43992 Onsala, Sweden
}

\pubyear{2019}
\volume{356}  
\jname{Nuclear Activity in Galaxies Across Cosmic Time}
\editors{M. Povic et~al.}
\begin{document}

\maketitle

\begin{abstract}
Active galactic nuclei (AGN) are believed to regulate star formation inside their host galaxies through ``AGN feedback". We summarise our on-going study of luminous AGN (z$\sim$0.2-3; $L_{\rm AGN,bol}$$\gtrsim$$10^{43}$\,erg\,s$^{-1}$), which is designed to search for observational signatures of feedback by combining observed star-formation rate (SFR) measurements from statistical samples with cosmological model predictions. Using the EAGLE hydrodynamical cosmological simulations, in combination with our $Herschel+$ALMA surveys, we show that - even in the presence of AGN feedback - we do not necessarily expect to see any relationships between average galaxy-wide SFRs and instantaneous AGN luminosities. We caution that the correlation with stellar mass for both SFR and AGN luminosity can contribute to apparent observed positive trends between these two quantities. On the other hand, the EAGLE simulations, which reproduce our observations, predict that a signature of AGN feedback can be seen in the wide specific SFR distributions of $all$ massive galaxies (not just AGN hosts). Overall, whilst we can not rule out that AGN have an immediate small-scale impact on in-situ star-formation, all of our results are consistent with a feedback model where galaxy-wide in-situ star formation is not rapidly suppressed by AGN, but where the feedback likely acts over a longer timescale than a single AGN episode. 
\end{abstract}

\firstsection 
\section{Introduction}
A fundamental component of galaxy formation models is that the central growing supermassive black holes (i.e., active galactic nuclei; AGN) regulate star formation inside their host galaxies. There have been considerable attempts to search for observational evidence of this ``AGN feedback'' throughout the last one-to-two decades (e.g., see review in \cite[Harrison 2017]{Harrison17}). One common approach has been to simultaneously measure AGN luminosities and the in-situ star-formation rates (SFRs) of AGN host galaxies. Here we refer to the in-situ SFRs as the measured {\em galaxy-wide} SFRs of on-going star-formation inside galaxies that host an observable AGN. In this article we summarise our own work following this approach, stressing the importance of testing specific model predictions. In a series of papers we have explored the SFRs and specific SFRs (sSFRs; SFR/stellar mass) of AGN-host galaxies (\cite[Stanley et~al. 2015, 2017, 2018]{Stanley15,Stanley17,Stanley18}; \cite[Harrison 2017]{Harrison17}; \cite[Scholtz et~al. 2018]{Scholtz18}). In these works we infer SFRs from the observed far-infrared luminosities after subtracting the AGN contribution using careful decomposition of the spectral energy distributions (following \cite[Stanley et~al. 2015]{Stanley15}). Removing the AGN contribution is important, particularly for the most powerful AGN, where the contribution to the total far-infrared luminosity can become significant (e.g., Fig.~7 of \cite[Stanley et~al. 2017]{Stanley17}). 

\section{AGN feedback does not reduce galaxy-wide in-situ SFRs}
  
\begin{figure}
\begin{center}
\includegraphics[width=0.45\textwidth,angle=90]{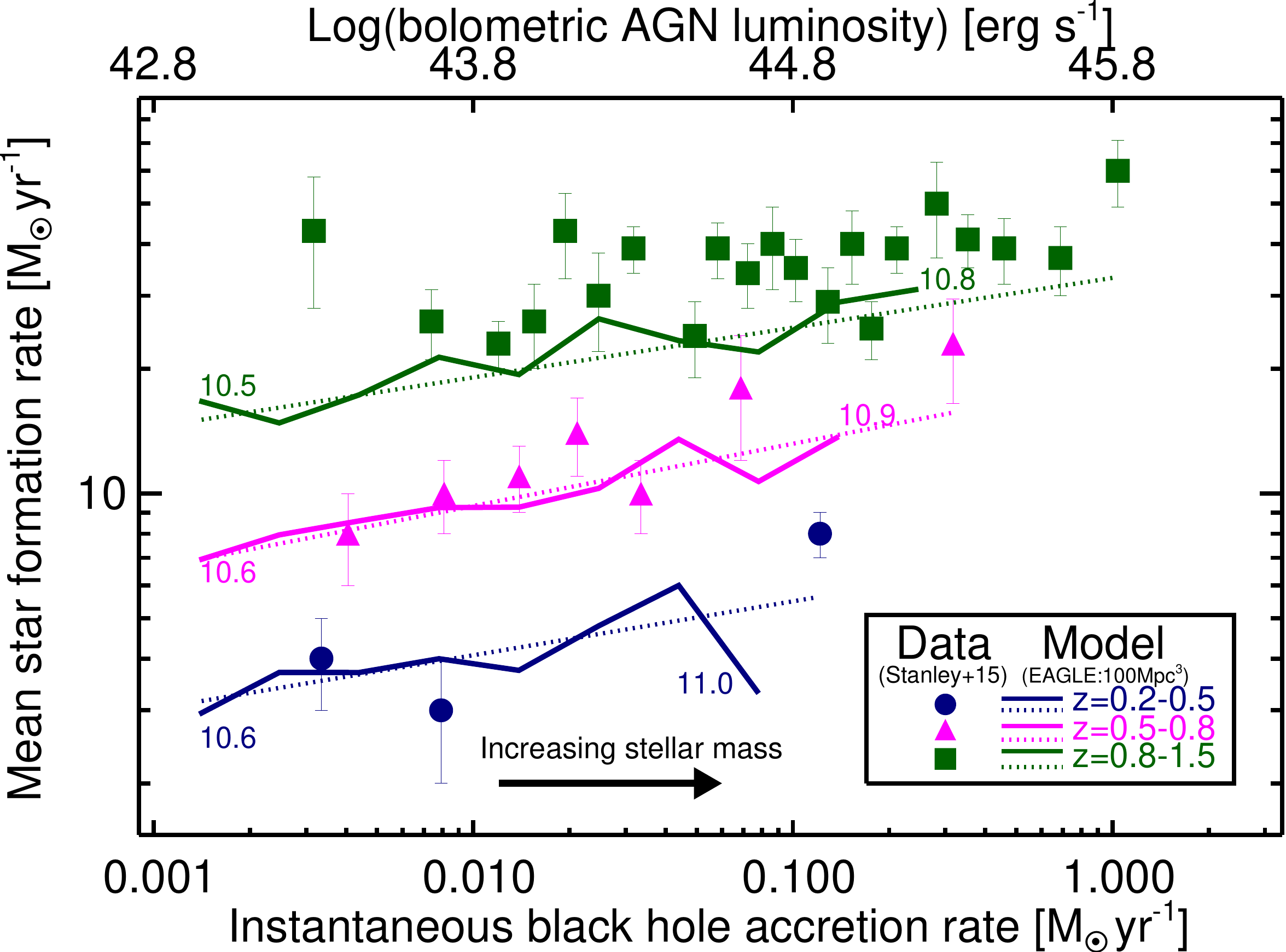}
 \caption{Mean star formation rate versus instantaneous
black hole accretion rate for the reference EAGLE simulation and versus AGN luminosity (converted from X-ray luminosity) from observations. The solid curves show the running average (mean) simulation values and the dotted lines are a linear fit to these values. The logarithm of the average stellar masses (in solar mass units) of the first and last values from the simulation are labelled. The slight increase in mean SFR with AGN luminosity is attributed to the increasing stellar masses. Effective star-formation suppression by AGN feedback does not infer that galaxies with a currently visible AGN should have reduced average {\em in-situ} galaxy-wide SFRs. Figure from \cite[Harrison (2017)]{Harrison17}.}
   \label{fig:Harrison17}
\end{center}
\end{figure}

\begin{figure}
\begin{center}
\includegraphics[width=0.65\textwidth,angle=0]{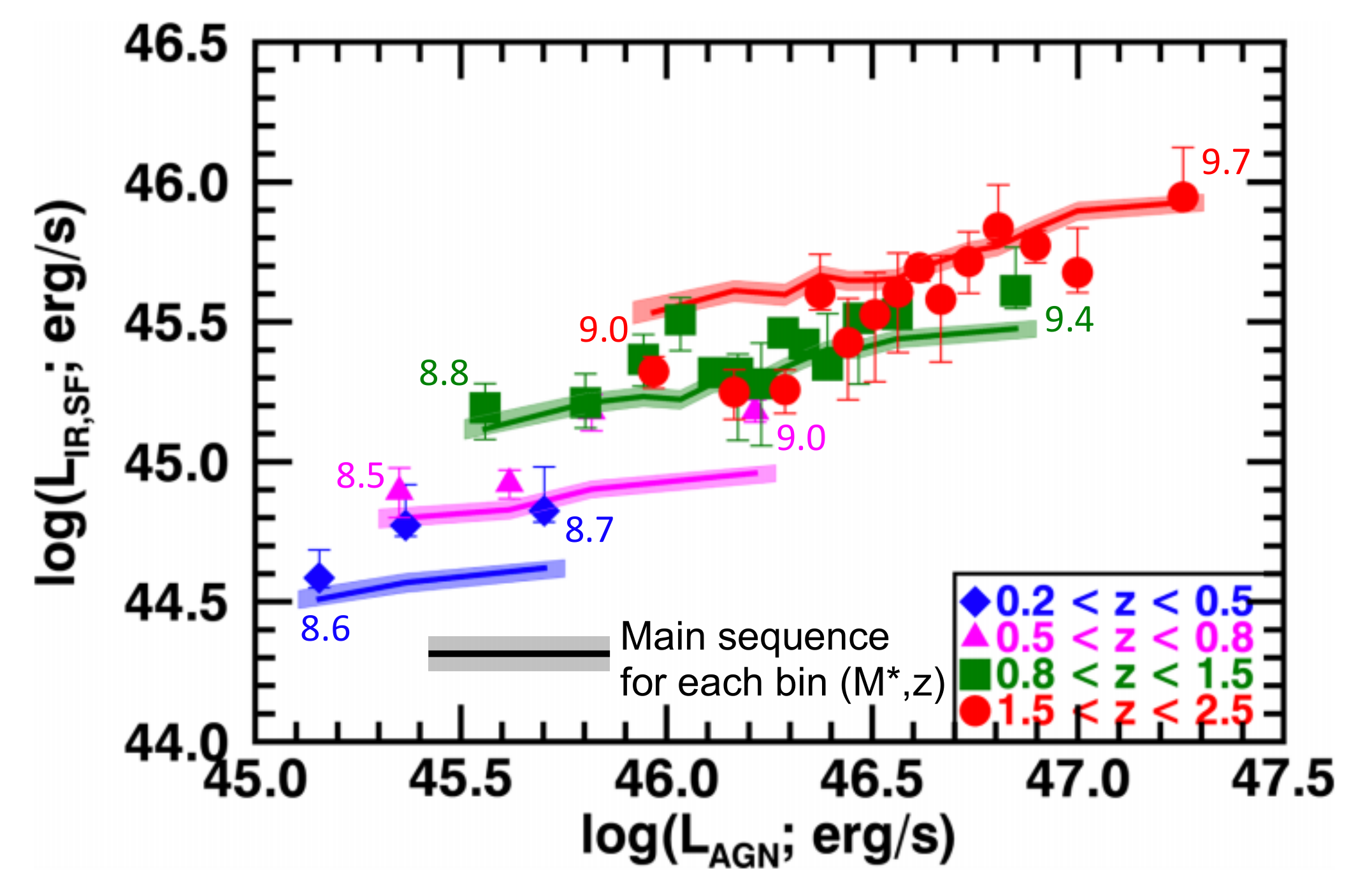}
 \caption{Mean infrared star formation luminosity (a proxy for SFR) versus bolometric AGN luminosity for Type~1 quasars. The solid curves show the expected value for star-forming galaxies at the average redshifts and stellar masses (extrapolated from the black hole masses) in each bin of quasars (\cite[Schreiber et al. 2015]{Scheiber15}). The logarithm of average black masses (solar mass units) of the first and last bins are labelled. Increasing mean SFR with increasing AGN luminosity is attributed to the increasing masses. Quasar host galaxies have {\em mean} SFRs that are consistent with mass- and redshift-matched star-forming galaxies; however, the underlying distributions may still be different (see Section~3 and Fig.~\ref{fig:Scholtz18sSFRs}). Figure adapted from \cite[Stanley et~al. (2017)]{Stanley17}.}
   \label{fig:Stanley17}
\end{center}
\end{figure}

In \cite[Stanley et~al. (2015)]{Stanley15} we combined {\em Herschel} and {\em Spitzer} photometry for $\sim$2000 $z$$=$$0.2$--2.5 X-ray AGN ($L_{\rm 2-10keV}=$10$^{42}$--10$^{45.5}$\,erg\,s$^{-1}$) and calculated average (mean) SFRs in bins of X-ray luminosity (data points in Fig.~\ref{fig:Harrison17}). Tracking the evolution of the overall galaxy population, a strong evolution of average SFR with redshift is observed. However, we find that the relationship between average SFR and AGN luminosity is only weakly correlated across all AGN luminosities investigated. In \cite[Stanley et~al. (2017)]{Stanley17} we repeated a similar experiment, over the same redshift range, but on powerful Type~1 quasars ($L_{\rm AGN}>10^{45}$\,erg\,s$^{-1}$) and using {\em WISE}$+${\em Herschel} data. We found a stronger correlation between SFR and AGN luminosity (Fig.~\ref{fig:Stanley17}); however, we showed that this can be explained by both quantities being correlated with stellar masses (assumed from black hole masses; see details in \cite[Stanley et~al. 2017]{Stanley17}). This mass effect is comparatively weak in the X-ray AGN sample in Fig.~\ref{fig:Harrison17}, which can be explained due to the different selection effects of X-ray AGN and optical Type~1 quasars, where the latter selects a narrower range of (higher) Eddington-ratios (see discussion in \cite[Rosario et~al. 2013]{Rosario13}). 

Across our work we found no strong dependence of average SFR on AGN luminosity when mass and redshift trends are accounted for. Furthermore, the {\em mean} SFRs of AGN-host galaxies are broadly consistent with star-forming ``main sequence'' galaxies (Fig.~\ref{fig:Stanley17}).
 
 In \cite[Harrison (2017)]{Harrison17}, to aid the interpretation of our observational results, we compared the observations of \cite[Stanley et~al. (2015)]{Stanley15} to the reference model of the EAGLE hydrodynamical cosmological simulations (\cite[Schaye et~al. 2015]{Schaye15}). The simulation models astrophysical processes (including prescriptions for star formation and AGN feedback) inside a 100\,Mpc$^{3}$ volume of the Universe and contains 1000s of massive galaxies, allowing us to track SFRs and AGN luminosities (inferred from instantaneous black hole accretion rates) across cosmic time. Consequently, we selected simulated galaxies following the same criteria as used for our observations (for details see \cite[McAlpine et~al. 2017]{McAlpine17}; \cite[Harrison 2017]{Harrison17}). EAGLE successfully reproduces the observed average SFRs on AGN (Fig.~\ref{fig:Harrison17}). This leads to an important conclusion: {\em no evidence for ``suppressed'' galaxy-wide in-situ SFRs in AGN host galaxies is still consistent with a cosmological model including AGN feedback}. This, perhaps counter intuitive result, can be explained if the timescale of an AGN episode is shorter than the timescale of star formation suppression by these (possibly multiple) AGN episode(s) (\cite[Harrison 2017]{Harrison17}; \cite[Scholtz et~al. 2018]{Scholtz18} and references there-in). 

\section{AGN feedback is imprinted on the overall galaxy population}

\begin{figure}
\begin{center}
\includegraphics[width=0.495\textwidth,angle=0]{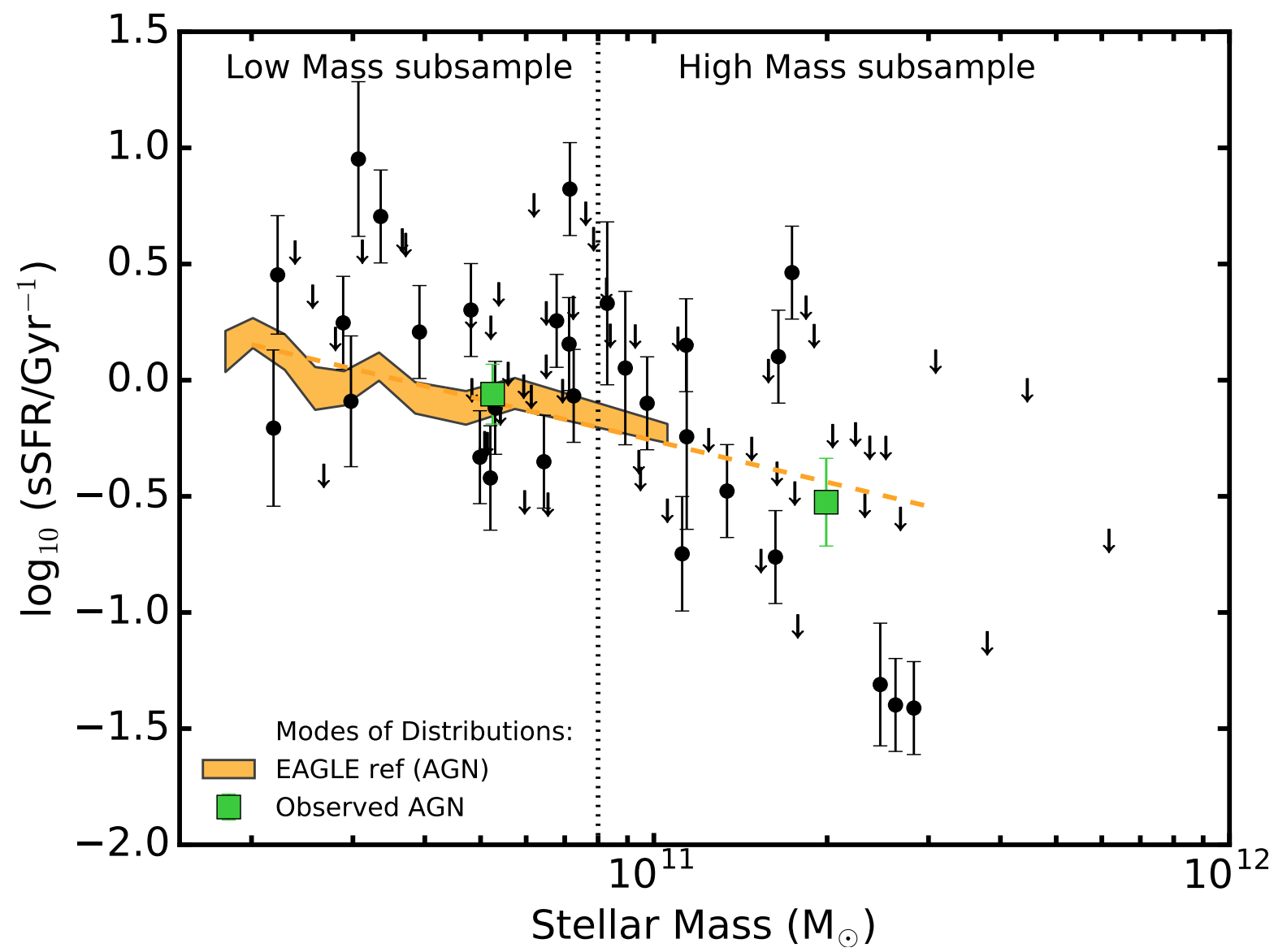}
\includegraphics[width=0.495\textwidth,angle=0]{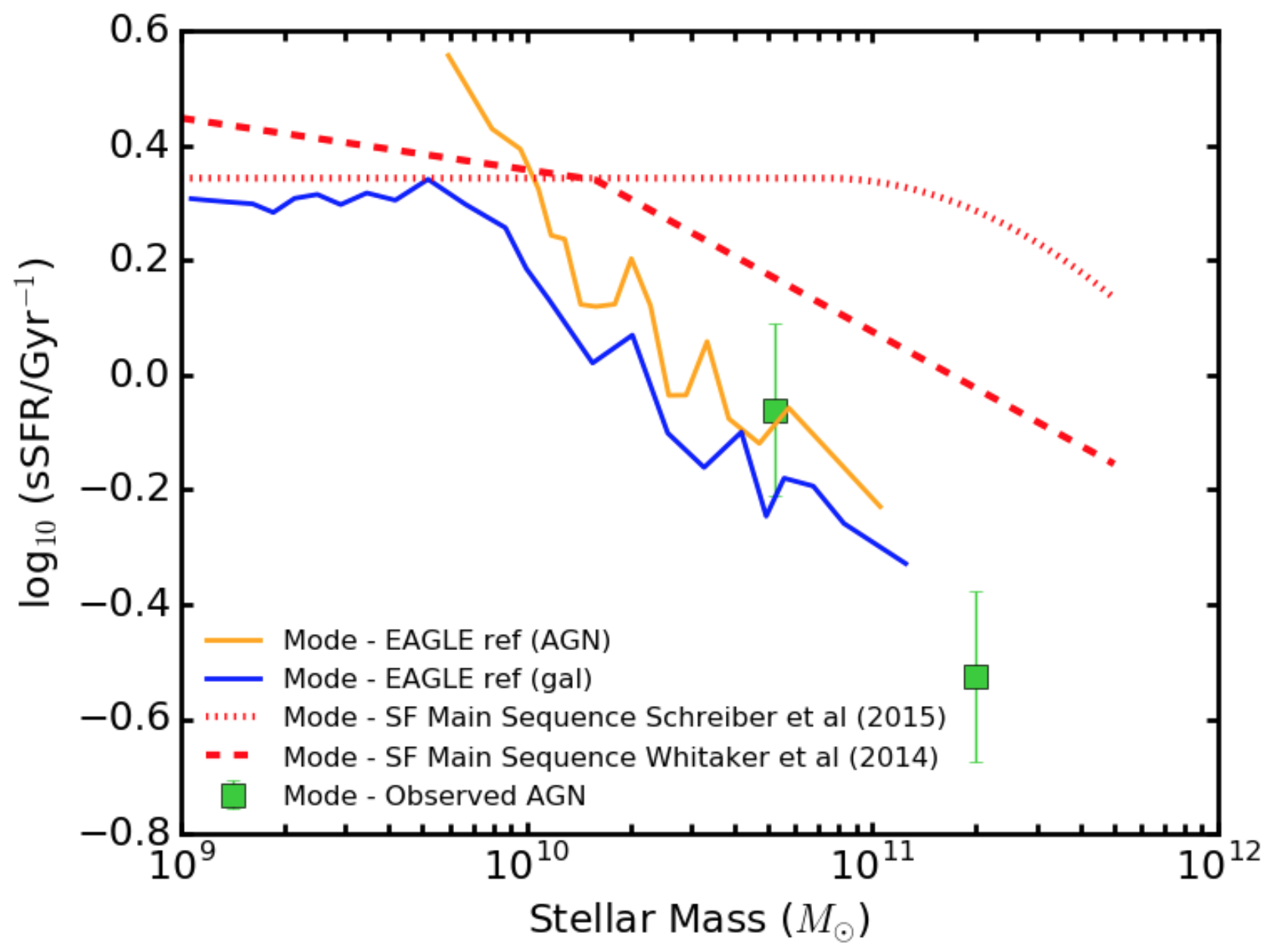}
 \caption{{\em Left:} sSFR versus stellar mass for X-ray AGN (black data points). Squares indicate the distribution mode for low and high mass subsamples (see vertical dashed line). The shaded region indicates the sSFR distribution mode for redshift-matched AGN host galaxies from EAGLE; the dashed line is an extrapolation to higher masses. {\em Right:} sSFR distribution modes versus stellar mass for X-ray AGN (data points) and for AGN hosts and all galaxies in the EAGLE reference model (blue and yellow curves, respectively). EAGLE matches the observations for the sSFRs for AGN hosts. However, AGN of these luminosities, have a lower {\em mode} of sSFRs than galaxies on the main sequence of {\em star-forming} galaxies (definitions shown from \cite[Schreiber et al. 2015]{Scheiber15} [dotted curve] and \cite[Whitaker et al. 2014]{Whitiker14} [dashed curve]). Figures from \cite[Scholtz et~al. (2018)]{Scholtz18}.}
   \label{fig:Scholtz18sSFRs}
\end{center}
\end{figure}

To move beyond simple average SFRs towards characterising distributions we obtained deep ALMA observations of $\sim$110 X-ray AGN, enabling us to achieve 2-10$\times$ more sensitive SFR constraints than previously possible (\cite[Stanley et~al. 2018]{Stanley18}). Using these results, in 
\cite[Scholtz et~al. (2018)]{Scholtz18}, we investigated $z$=1.5--3.2 X-ray AGN ($L_{\rm 2-10keV}=$10$^{43}$--10$^{45}$\,erg\,s$^{-1}$), and measured the mode and width of the (specific) SFR distributions of AGN host galaxies (following \cite[Mullaney et~al. 2015]{Mullaney15}). We found that, for AGN in this luminosity range, whilst the {\em mean} SFRs of AGN are typically consistent with the main sequence of star-forming galaxies, the {\em median/mode} SFRs of AGN host galaxies are typically found to be {\em lower} than star-forming galaxies (Fig.~\ref{fig:Scholtz18sSFRs}). This can be explained because of different underlying distributions, where AGN host galaxies have a broader distribution, extending to lower SFRs compared to the main sequence (\cite[Mullaney et~al. 2015]{Mullaney15}).\footnote{We note, with increasing AGN luminosity there may be an increased likelihood for AGN-host galaxies to be star-forming main sequence galaxies (\cite[Bernhard et~al. 2019; Schulze et~al. 2019]{Bernhard19,Schulze19}).} Nonetheless, for feedback studies we suggest that it is more meaningful to investigate the galaxy population {\em as a whole} instead of a comparison to the main sequence (which is still not well defined for the highest masses; see dotted/dashed curves in Fig.~\ref{fig:Scholtz18sSFRs}).

\begin{figure}[!h]
\begin{center}
\includegraphics[width=0.7\textwidth]{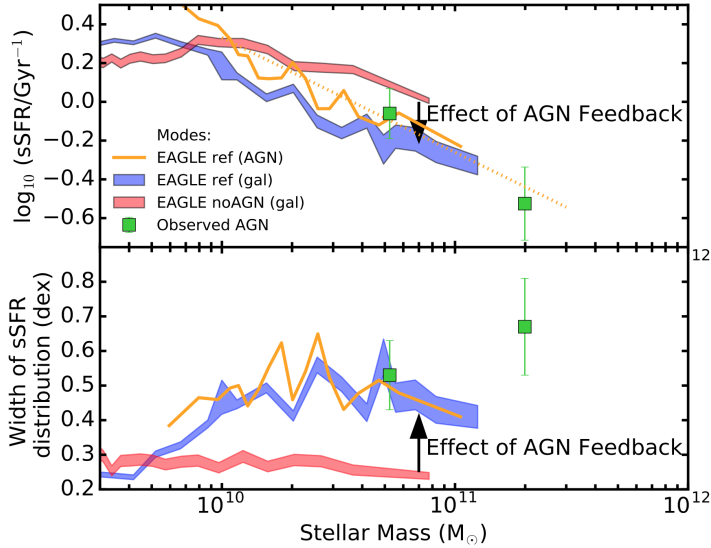}
 \caption{Mode (top panel) and width (bottom panel) of sSFR distributions versus stellar mass for observed AGN hosts (data points) compared to {\em all} galaxies in both the EAGLE reference simulation and the EAGLE simulation without AGN (blue and red shaded regions, respectively). The orange solid curve is for AGN host galaxies in the reference model (extrapolated to higher masses with the dashed line). In the EAGLE reference simulation (which matches the trends of the data), AGN feedback reduces the mode and increases the width of sSFR distributions. This is seen in the distributions of {\em all} massive galaxies, even if the AGN happen to be ``off'' at the time that they are observed. Figure from \cite[Scholtz et~al. (2018)]{Scholtz18}.}
   \label{fig:Scholtz18}
\end{center}
\end{figure}

In \cite[Scholtz et~al. (2018)]{Scholtz18}, we investigated galaxies in the EAGLE simulations with the AGN turned off (i.e., with no AGN feedback). This enabled us to show that, within the main reference model, the impact of AGN is to decrease the mode, by a factor of $\sim$2--3, and to increase the width, by a factor of $\sim$2--3, of the sSFR distributions of massive galaxies (Fig.~\ref{fig:Scholtz18}). This is simply explained by the fact that AGN feedback produces the quiescent galaxy population, spreading the sSFR to lower values (\cite[Scholtz et~al. 2018]{Scholtz18}).

At the limit of the spatial resolution of our integral field spectroscopic observations ($\sim$few kpc), in combination with high-resolution far-infrared imaging, we have also found no evidence that AGN-driven ionised outflows (traced via high-velocity [O~{\sc iii}] emission) have an instantaneous positive or negative significant impact on the in-situ star-formation in a representative sample of 8 $z$$\sim$$1.4$--2.6 AGN (\cite[Scholtz et~al. 2020]{Scholtz19}). This is in qualitative agreement with at least some models that suggest that AGN outflows have no impact upon the {\em in-situ} star formation (\cite[Gabor \& Bournaud 2014]{Gabor14}). Nonetheless, we can not rule out a smaller-scale impact or that outflows have an impact on longer timescales; for example by removing gas which is later prevented from re-accreting onto the galaxy.

We have shown that if AGN feedback is the mechanism to suppress galaxy-wide star formation in massive galaxies it does not necessarily follow that AGN host galaxies have ``suppressed'' (specific) SFRs compared to, mass- and redshift-matched, galaxies without a visible AGN (Fig.~\ref{fig:Harrison17}). Instead, the signature of AGN feedback is likely to be imprinted on the properties of the entire massive galaxy population (Fig.~\ref{fig:Scholtz18}). More work is now required to test specific observable model predictions that use different prescriptions for AGN feedback. For example, assessing if the observed molecular gas content of AGN host galaxies (e.g., \cite[Shangguan et~al. 2018]{Shangguan18}) is consistent with model predictions.  

\section{Final remarks: the need for testing specific model predictions}
We have investigated the SFRs of large samples of AGN host galaxies. We have emphasised the importance of controlling for redshift and mass when investigating the SFRs of AGN host galaxies - without doing so could result in artificial correlations between SFRs and AGN luminosities. By comparing our observations with the EAGLE simulations we have concluded that: (1) AGN feedback does not reduce galaxy-wide in-situ SFRs of AGN host galaxies; (2) the signature of AGN feedback is imprinted on the overall massive galaxy population. Importantly, any observation which finds that galaxy-wide SFRs of AGN host galaxies are not ``suppressed'' does not rule out all models of AGN feedback. Instead this observation may only rule out models where the complete suppression of star formation is more rapid than the AGN episode itself. {\em When looking for observational signatures of AGN feedback, we strongly advocate testing specific model predictions rather than just expecting AGN host galaxies to be special.} Further rigorous model--observation comparisons are required to make progress in understanding how AGN impact upon the star formation in their host galaxies.

\end{document}